\documentclass{article}
\usepackage{latexsym,amssymb,amsmath,amscd,amsthm,amsxtra}
\usepackage{amsmath}
\usepackage{amsfonts}
\usepackage{amssymb}
\usepackage{graphicx}%
\usepackage[margin=1.5in]{geometry}
\newcommand{\gmconn}{\mbox {\boldmath $\partial$}}
\newcommand{\complex}{{\mathbb C}}

\newcommand{\reals}{{\mathbb R}}

\newcommand{\integers}{{\mathbb Z}}

\newcommand{\calb}{{\cal B}}

\newcommand{\cale}{{\cal E}}

\newcommand{\calh}{{\cal H}}

\newcommand{\call}{{\cal L}}
\newcommand{\calm}{{\cal M}}

\newcommand{\arrows}{\,\lower1pt\hbox{$\longrightarrow$}\hskip-.24in\raise2pt

              \hbox{$\longrightarrow$}\,}

\title{Quantization and holomorphic anomaly}
\author{Albert Schwarz\thanks{partly supported by NSF grant No. DMS 0505735.}\ \ and Xiang Tang\thanks{partly supported by NSF grant No. DMS
0505735 and 0604552.}}
\begin{document}
\maketitle
\begin{abstract}
We study wave functions of B-model on a Calabi-Yau threefold in
various polarizations.
\end{abstract}
\section{Introduction}

In present paper, we consider wave functions of B-model on a
Calabi-Yau threefold in various polarizations and  relations
between these wave functions.

One can interpret genus 0 B-model on a Calabi-Yau threefold $X$ as
a theory of variations of complex structures. The extended moduli
space of complex structures, the space of pairs (complex
structure, holomorphic 3-form), can be embedded into the
middle-dimensional cohomology $H^3(X, \complex)$ as a lagrangian
submanifold. The B-model for arbitrary genus coupled to gravity
(the B-model topological string) can be obtained from genus 0
B-model by means of quantization; the role of Planck constant is
played by $\lambda^2$ where $\lambda$ is the string coupling
constant. (This is a general statement valid for any topological
string. It was derived by Witten \cite{witten:hol-anomaly} from
worldsheet calculation of \cite{BCOV:hol-anomaly}. ) The partition
function of B-model is represented by wave function depending on
choices of polarization in $H^3(X, \complex)$. If the polarization
does not depend holomorphically on the points of the moduli space
of complex structures, then the dependence of wave function of the
points of the moduli space is not necessarily holomorphic. (The
$\bar{t}$-dependence is governed by the holomorphic anomaly
equation.) This happens, in particular, for a polarization that we
call  complex hermitian polarization. Other papers use the term
``holomorphic polarization" for a complex hermitian polarization
in the sense of present paper; we reserve the term ``holomorphic
polarization" for a polarization that depends holomorphically on
the points of the moduli space of complex structures. The
holomorphic polarization in our sense was widely used in mirror
symmetry; this polarization and its $p$-adic analog were used to
analyze integrality of instanton numbers (genus 0 Gopakumar-Vafa
invariants) \cite{ksv:integrality}.

The main goal of present paper is to study wave functions in
various polarizations, especially in holomorphic polarization. We
believe that Gopakumar-Vafa invariants for any genus can be
defined by means of $p$-adic methods and this definition will have
as a consequence integrality of these invariants. The present
paper is a necessary  first step in the realization of  this
program. It served as a basis for a conjecture about the structure
of Frobenius map on $p$-adic wave functions formulated in \cite
{sv}; this conjecture implies integrality of Gopakumar-Vafa
invariants.

We begin with a short review of quantization  of symplectic vector
space (Section 2).  In Section 3, 4 and 5 we use the general
results of Section 2 to obtain relations between the wave
functions of B-model in real, complex hermitian and holomorphic
polarizations. In Section 6 we compare these wave functions with
worldsheet calculations of \cite{BCOV:hol-anomaly}.

The holomorphic anomaly equations were recently studied and
applied in [1], [5], [6], [9], [11]. Some of equations in our
paper differ slightly from corresponding equations in [1], [11].
However, this difference does not affect  any conclusions of these
papers.

{\bf Acknowledgments:} We would like to thank M. Aganagic and M.
Kontsevich for helpful discussions.
\section{Quantization}

We consider a real symplectic vector space $V$ and a symplectic
basis of $V$. (A symplectic structure can be considered as a skew
symmetric non-degenerate bilinear form $<\ ,\ >$ on $V$; we say
that $e^\alpha,\ e_\beta$, $\alpha, \beta=1, \cdots, n={\text
{dim}}(V)$ is a symplectic basis if $<e^\alpha,
e^\beta>=<e_\alpha, e_\beta>=0$, $<e^\alpha,
e_\beta>=\delta_\beta^\alpha$.) It is well known that for every
symplectic basis $e=\{e^\alpha, e_\beta\}$, one can construct a
Hilbert space $\calh_e$; these spaces form a bundle over the space
$\calm$ of all bases and one can construct a projectively flat
connection on this bundle\footnote{One says that a connection
$\nabla$ on a vector bundle over a space $B$ is projectively flat
if $[\nabla_X, \nabla_Y]=\nabla_{[X,Y]}+C$, where $C$ is a
constant depending on $X, Y$. For an infinite dimensional vector
bundle over a compact manifold $B$ with a unitary connection, this
means that for every two points $e, \tilde{e}$ of $B$ connected by
a continuous path in $B$, there is an isomorphism between the
fibers $\calh_e$ and $\calh_{\tilde{e}}$ defined up to
multiplication by a constant; this isomorphism depends on the
homotopy class of the path. We say that a section $\Phi $ of the
vector bundle is projectively flat if $\nabla _X \Phi =C_X \Phi$
where $C_X$ is a scalar function on the base.}. The situation does
not change if we consider, instead of a real basis in $V$, a basis
$\{e^\alpha, e_\alpha\}$ in the complexification of $V$ requiring
that $e_\alpha$ be complex conjugate to $e^\alpha$.

The picture we described above is the standard picture of
quantization of a symplectic vector space. The choice of a basis in
$V$ specifies a real polarization; the choice of a basis in its
complexification determines a complex polarization. The quantum
mechanics lives in Hilbert space of functions depending on
$n=\frac{1}{2}\text{dim} V$ variables. To construct this Hilbert
space, we should fix a polarization, but Hilbert spaces
corresponding to different polarizations can be identified up to a
constant factor. In semiclassical approximation, vectors in
Hilbert space correspond to lagrangian submanifolds of $V$.

Let us describe the Hilbert space $\calh_e$ for the case when
$e=\{e^\alpha, e_\beta\}$ is a symplectic basis of $V$. An element
of $V$ can be represented as a linear combination of vectors
$e^\alpha, e_\beta$ with coefficients $x_\alpha, x^\beta$. After
quantization, $x_\alpha$ and $x^\beta$ become self-adjoint
operators $\hat{x}_\alpha$, $\hat{x}^\beta$ obeying canonical
commutation relations({\bf CCR}):
\begin{equation}
\label{eq:ccr} [\hat{x}_\alpha, \hat{x}_\beta]=[\hat{x}^\alpha,
\hat{x}^\beta]=0,\ \ \ \ [\hat{x}_\alpha,
\hat{x}^\beta]=\frac{\hbar}{i}\delta_\alpha^\beta
\end{equation}
We define $\calh_e$ as the space of irreducible unitary
representation of canonical commutation relations.

A (linear) symplectic transformation transforms a symplectic basis
$\{e^\alpha, e_\alpha\}$ into symplectic basis
$\{\tilde{e}^\alpha, \tilde{e}_\alpha\}$:
\begin{equation}
\begin{array}{ll}
\label{eq:symp-basis-trans}\tilde{e}^\alpha&=M^\alpha_\beta
e^\beta+N^{\alpha\beta}e_\beta\\
\tilde{e}_\alpha&=R_{\alpha\beta}e^\beta+S_\alpha^\beta e_\beta.
\end{array}
\end{equation}
This transformation acts on $\hat{x}_\alpha, \hat{x}^\alpha$ as a
canonical transformation, i.e. the new operators
$\hat{\tilde{x}}_\alpha, \hat{\tilde{x}}^\alpha$ also obey {\bf
CCR}; they are related to $\hat{x}_\alpha, \hat{x}^\alpha$ by the
formula:
\begin{equation}
\begin{array}{ll}
\hat{x}^\alpha&=N^{\beta\alpha}\hat{\tilde{x}}_\beta+S^\alpha_\beta
\hat{\tilde{x}}^\beta\\
\hat{x}_\alpha&=M_\alpha^\beta
\hat{\tilde{x}}_\beta+R_{\beta\alpha}\hat{\tilde{x}}^\beta.
\end{array}
\end{equation}

It follows from the uniqueness of unitary irreducible
representation of {\bf CCR} that there exists a unitary operator
$T$ obeying
\begin{equation}
\label{eq:trans-T}
\begin{array}{ll}
\hat{\tilde{x}}^\alpha&=T\hat{x}^\alpha T^{-1},\\
\hat{\tilde{x}}_\alpha&=T\hat{x}_\alpha T^{-1}.
\end{array}
\end{equation}
This operator $T$ is defined up to a constant factor relating
$\calh_e$ and $\calh_{\tilde{e}}$. In the case when
$\{\tilde{e}^\alpha, \tilde{e}_\alpha\}$ is an infinitesimal
variation of $\{e^\alpha, e_\alpha\}$, i.e. $\tilde{e}=e+\delta e$
where
\begin{equation}
\begin{array}{ll}
\delta e^\alpha&=m_\beta^\alpha e^\beta+n^{\alpha\beta}e_\beta,\\
\delta e_\alpha&=r_{\alpha\beta}e^\beta+s_\alpha ^\beta e_\beta,
\end{array}
\end{equation}
we can represent the operator $T$ as $1+\delta T$, where
\begin{equation}
\delta T=-\frac{1}{2\hbar} n^{\alpha\beta}\hat{x}_\alpha
\hat{x}_\beta+\frac{1}{\hbar}m_\alpha^\beta \hat{x}^\alpha
\hat{x}_\beta-\frac{1}{2\hbar}r_{\alpha\beta}\hat{x}^\alpha\hat{x}^\beta+C.
\end{equation}
This formula determines a projectively flat connection on the
bundle with fibers $\calh_e$ and the base consisting of all
symplectic bases in $V$. A quantum state specifies a projectively
flat section of this bundle.

The irreducible unitary representation of {\bf CCR} can be
realized by operators of multiplication and differentiation on the
space of square integrable functions of $x^1, \cdots, x^n$; one
can take $\hat{x}^\alpha \Psi=x^\alpha \Psi$ and
$\hat{x}_\alpha\Psi=\frac{\hbar}{i}\frac{\partial \Psi}{\partial
x^\alpha}$. Then a projectively flat connection takes the form:
\begin{equation}\label{eq:quant-trans}
\delta\Psi=\frac{\hbar}{2}n^{\alpha\beta}\frac{\partial^2
\Psi}{\partial x^\alpha\partial x^\beta}+m^\beta_\alpha x^\alpha
\frac{\partial \Psi}{\partial
x^\beta}-\frac{1}{2\hbar}r_{\alpha\beta}x^\alpha x^\beta
\Psi+C\Psi.
\end{equation}

We will call elements of $\calh_e$ and corresponding functions of
$x^1, \cdots, x^n$ wave functions.

It is important to notice that in Equation (\ref{eq:quant-trans})
instead of square integrable functions, we can consider functions
$\Psi(x^1, \cdots, x^n)$ from an almost arbitrary space $\cale$; the
only essential requirement is that the multiplication by
$x^\alpha$ and differentiation with respect to $x^\alpha$ should
be defined on a dense subset of $\cale$ and transform this set
into itself.

Sometimes it is convenient  to restrict ourselves to the space of
functions of the form $\Psi=\exp (\frac{\Phi}{\hbar})$ where $\Phi
=\sum \varphi _n \hbar ^n$ is a formal series with respect to
$\hbar$ (semiclassical wave functions). Rewriting (\ref
{eq:quant-trans}) on this space we obtain
$$
\delta \Phi =\frac{1}{2}n^{\alpha\beta}(\hbar\frac{\partial^2
\Phi} {\partial x^\alpha\partial x^\beta}+\frac{\partial
\Phi}{\partial x^\alpha}\frac{\partial \Phi}{\partial
x^\beta})+m_\alpha^\beta x^\alpha \frac{\partial \Phi}{\partial
x^\beta}-\frac{1}{2}r_{\alpha\beta}x^\alpha x^\beta+\hbar C.$$

Let $B$ be the set of all symplectic bases in the complexification
of $V$. We consider the total space of a bundle over $B$ as the
direct product $B\times \cale$. One can use Equation
(\ref{eq:quant-trans}) to define a projectively flat connection on
this vector bundle. (The coefficients of infinitesimal variation
(\ref{eq:symp-basis-trans}) of the basis in $V$ must be real; if
we consider $\{e^\alpha, e_\alpha \}$ as a basis of
complexification of $V$, the coefficients of infinitesimal
variation obey the same conditions
$n^{\alpha\beta}=n^{\beta\alpha}, r_{\alpha\beta}=r_{\beta\alpha},
m^\alpha_\beta+s^\beta_\alpha=0$, but they can be complex.)

Notice, however that in the real case we are dealing with unitary
connection; the operator $T_{e, \tilde{e}}$ that identifies two
fibers (up to a constant factor) always exists. In complex case,
the equation for projectively flat section can have solutions only
over a part of the set of symplectic bases. (Recall that the
fibers of our vector bundles are infinite-dimensional.)

It is easy to write down simple formulas for the operator $T_{e,
\tilde{e}}$ in the case when $N^{\alpha\beta}=0$ or
$R_{\alpha\beta}=0$. In the first case we have
\begin{equation}
\label{eq:r}
T_{e,\tilde{e}}(\Psi)(x^\alpha)=\exp\big(-{\frac{1}{2\hbar}(RM^{-1})
_{\alpha\gamma}x^\alpha x^\gamma}\big) \Psi(S^\alpha_\beta
x^\beta),
\end{equation}
in the second case
\begin{equation}
\label{eq:n}
T_{e,\tilde{e}}(\Psi)(x^\alpha)=\exp\big(-\frac{\hbar}{2}(MN^{T})^{\alpha\gamma}
\frac{\partial^2}{\partial x^\alpha\partial
x^{\gamma}}\big)\Psi(S^\alpha_\beta x^\beta).
\end{equation}
Combining Equations (\ref{eq:r}) and (\ref{eq:n}), we obtain an
expression for $T_{e, \tilde{e}}$ that is valid when $M$ and $S$
are non-degenerated matrices,
\begin{equation}
\label{eq:T}
T_{e,\tilde{e}}\Psi(x^\alpha)=\exp\big(-\frac{1}{2\hbar
}(RM^{-1})_{\alpha\gamma}x^\alpha
x^\gamma\big)\big\{\exp\big(-\frac{\hbar}{2}(MN^T)^{\alpha\gamma}
\frac{\partial^2}{\partial x^\alpha\partial
x^\gamma}\big)\big[\Psi((M^{-1})^\alpha_\beta x^\beta)\big]\big\}.
\end{equation}

Using the expression (\ref{eq:T}) and Wick's theorem, it is easy
to construct diagram techniques to calculate $T_{e,
\tilde{e}}e^{F}$.

Recall that  Wick's theorem permits us to represent an expression
of the form $\int e^Ae^{V(x)} dx$ where $A$ is a quadratic form and $V(x)$ does not contain linear and quadratic terms as a sum  of Feynman diagrams :
\begin{equation}
\label{ }
\int \exp (\frac{1}{2}a_{ij}{
x^i x^j})e^{V(x)}dx=e^{W}
\end{equation}
where $W$ is a sum of connected Feynman diagrams
with propagator $a^{ij}$ (inverse to $a_{ij}$)  and with vertices determined by $V(x)$. Using this fact 
and Fourier transform we obtain a diagram technique for 
$T_{e,
\tilde{e}}e^{F}$.

It follows from this statement  that the action of $T_{e,\tilde{e}}$ on the space of semiclassical wave functions is given by rational expressions. This means that the action can be defined over an arbitrary field (in particular, over $p$-adic numbers).

The above statements can be reformulated in the language of
representation theory. Assigning to every symplectic
transformation (\ref{eq:trans-T}) a unitary operator $T$ defined
by $(\ref{eq:quant-trans})$ we obtain a multivalued representation
of the symplectic group $Sp(n, \reals)$  and corresponding Lie
algebra $sp(n, \reals)$ (metaplectic representation). The
representation of the Lie algebra $sp(n, \reals)$ can be extended
to a representation of its complexification $sp(n, \complex)$ in
an obvious way. However, the metaplectic representation of $Sp(n,
\reals)$ cannot be extended to a representation of $Sp(n,
\complex)$ because $Sp(n, \complex)$ is simply connected and
therefore it does not have any non-trivial multivalued
representations. (See Deligne \cite{de:metaplectic} for more
detailed analysis.)

\section{B-model}
From the mathematical viewpoint, the genus 0 B-model on a compact
Calabi-Yau threefold $X$ is a theory of variations of complex
structures on $X$. Let us denote by $\calm$ the moduli space of
complex structures on $X$. For every complex structure, we have a
non-vanishing holomorphic $(3,0)$-form $\Omega$ on $X$, defined up
to a constant factor. Assigning the set of forms $\lambda \Omega$
to every complex structure we obtain  a line bundle $\call$ over
$\calm$. The total space of this bundle, i.e. the space of all
pairs (complex structure on $X$,  form $\lambda \Omega$), will be
denoted by $\widetilde{\calm}$. Every form $\Omega$ specifies an
element of $H^3(X, \complex)$ (middle-dimensional cohomology of
$X$) that will be denoted by the same symbol. Notice that $\Omega$
depends on the complex structure on $X$, but $H^3(X, \complex)$
does not depend on complex structure. More precisely, the groups
$H^3(X, \complex)$ form a vector bundle over $\calm$ and this
bundle is equipped with a flat connection $\gmconn_a$ (Gauss-Manin
connection).  In other words, the groups $H^3(X, \complex)$ where
$X$ runs over small open subset of $\calm$ are canonically
isomorphic. However, the bundle at hand is not necessarily
trivial: the Gauss-Manin connection can have non-trivial
monodromies. Going around a closed homotopically non-trivial loop
$\gamma$ in $\calm$, we obtain a (possibly) non-trivial
isomorphism $M_\gamma: H^3(X, \complex)\to H^3(X, \complex)$. The
set of all elements of $H^3(X, \complex)$ corresponding to forms
$\Omega$ constitutes a lagrangian submanifold $L$ of $H^3(X,
\complex)$. (The cup product on $H^3(X, \complex)$ taking values
in $H^6(X, \complex)=\complex$ specifies a symplectic structure on
$H^3(X, \complex)$. The fact that $L$ is lagrangian follows
immediately from the Griffiths transversality.)  We can also say
that we have a family of lagrangian submanifolds $L_\tau\subset
H^3(X_\tau, \complex)$ where $H^3(X_\tau, \complex)$ denotes the
third cohomology of the manifold $X$ equipped with the complex
structure $\tau\in \calm$. Notice that the Lagrangian submanifold
$L$ is invariant with respect to the monodromy group (the group of
monodromy transformations $M_\gamma$).

The B-model on $X$ for an arbitrary genus can be obtained by means of
quantization of genus 0 theory, the role of the Planck constant is
played by $\lambda^2$, where $\lambda$ is the string coupling
constant. (More precisely, we should talk about $B$-model coupled
to gravity or about $B$-model topological string.) Let us fix a
symplectic basis $\{e_A, e^A\}$ in the vector space $H^3(X_\tau,
\complex)$. Every element $\omega \in H^3(X_\tau, \complex)$ can
be represented in the form $\omega=x^Ae_A+x_Ae^A$, where the
coordinates $x^A, x_A$ can be represented as $x^A=<e^A, \omega>,
x_A=-<e_A, \omega>$. Quantizing the symplectic vector space
$H^3(X_\tau, \complex)$ by means of polarization $\{e_A, e^A\}$,
we obtain a vector bundle $\calh$ with fibers $\calh_e$. (It would
be more precise to denote the fiber by $\calh_{\tau, e}$ stressing
that a point of the base of the bundle $\calh$ is a pair $(\tau,
e)$ where $\tau\in \calm$ and $e$ is a symplectic basis in
$H^3(X_\tau, \complex)$, however, we will use the notation
$\calh_e$, having in mind that the notation $e$ for the basis
already includes the information about the corresponding point
$\tau=\tau(e)$ of the moduli space $\calm$.) As usual, we have a
projectively flat connection on the bundle $\calh$. Let us denote
by $\calb$ the space of all symplectic bases in the cohomology
$H^3(X_\tau, \complex)$ where $\tau$ runs over the moduli space
$\calm$. Then the total space of the bundle $\calh$ can be
identified with the direct product $\calb\times \cale$, where
$\cale$ stands for the space of functions depending on $x^A$. Let
us suppose that the basis $\{e_A, e^A\}$ depends on the parameters
$\sigma^1, \cdots, \sigma^K$ and
\begin{equation}
\begin{array}{ll}
\partial_i e^A&=m_B^A e^B+n^{AB}e_B\\
\partial_i e_A&=r_{AB}e^B+s_A^Be_B,
\end{array}
\end{equation}
where in the calculation of the derivatives
$\partial_i=\frac{\partial}{\partial \sigma^i}$, we identify the
fibers $\calh_e$ by means of Gauss-Manin connection. Then a
projectively flat section $\Psi(x^A,\sigma^i, \lambda)$ satisfies
the following equation
\begin{equation}\label{eq:Z}
\frac{\partial \Psi}{\partial
\sigma^i}=\big[-\frac{1}{2}\lambda^{-2}r_{AB}x^A x^B+m_{B}^A
x^B\frac{\partial}{\partial x^A}+\frac{1}{2}\lambda^2
n^{BC}\frac{\partial^2}{\partial x^B
\partial x^C}+C_i(\sigma)\big]\Psi(x^A, \sigma^i, \lambda).
\end{equation}
This follows immediately from Equation (\ref{eq:quant-trans}).
(Recall that the wave function $\Psi$ depends on  half of
coordinates on the symplectic basis $\{e_A, e^A\}$.)

The wave function of the $B$-model topological string is a
projectively flat section $\Psi$ of the bundle $\calh$ that in
semiclassical approximation corresponds to the lagrangian
submanifold $L$ coming from genus 0 theory. Of course, such a
section is not unique and one needs additional assumptions to
determine the wave function.

Notice that the right object to consider in $B$-model is the wave
function $\Psi(x,e, \lambda)$ defined as a function of $x^A$ and
polarization $e=\{e_A, e^A\}$. However, it is convenient to work
with $\Psi$ restricted to certain subset of the set of
polarizations. In particular, we can fix an integral basis
$\{g_A(\tau), g^A(\tau)\}$ in $H^3(X_\tau, \complex)$ that varies
continuously with $\tau \in \calm$. (The integral vectors of
$H^3(X_\tau, \complex)$ are defined as vectors in the image of
integral cohomology $H^3(X_\tau, \integers)$ in $H^3(X_\tau,
\complex)$.) It is
obvious that the vectors $\{g_A, g^A\}$ are covariantly constant
with respect the Gauss-Manin connection, therefore we can assume
that in this polarization the wave function does not depend on the
point of moduli space. It can be represented in the form
\begin{equation}
\Psi_{\text{real}}(x^A, \lambda)=\exp\big[ \sum_{g=0}^\infty
\lambda^{2g-2}{\cal F}_g(x^A)\big],
\end{equation}
where ${\cal F}_g$ is the contribution of genus $g$ surfaces. The
leading term in the exponential as always specifies the
semiclassical approximation; it corresponds to the genus zero free
energy $F_0(x^A)$.
In the next section, we will calculate the transformation of the
wave function $\Psi$ from the real polarization to some other
polarizations.
It is important to emphasize that the Gauss-Manin connection can
have non-trivial monodromies, hence the integral basis
$\{g_A(\tau), g^A(\tau)\}$ is a multivalued function of $\tau\in
\calm$. The quantum state represented by the wave function
$\Psi_{\text{real}}$ should be invariant with respect to the monodromy
transformation $M_\gamma$; in other words, one can find such
numbers $c_\gamma$ that
\begin{equation}\label{eq:monodromy}
\widetilde{M}_\gamma\Psi_{\text{real}}=c_\gamma\Psi_{\text{real}},
\end{equation}
where $\widetilde{M}_\gamma$ stands for the transformation of wave
function corresponding to the symplectic transformation $M_\gamma$
(in other words $\widetilde{M}_\gamma$ corresponds to $M_\gamma$
under metaplectic representation). The condition
(\ref{eq:monodromy}) imposes severe restrictions on the state $\Psi_{\text{real}}$, but it does not determine $\Psi_{\text{real}}$
uniquely.

In the next section, we will calculate the transformation of the
wave function $\Psi$ from the real polarization to some other
polarizations.

\section{Complex hermitian polarization}
Let us introduce special coordinates on $\calm$ and
$\widetilde{\calm}$. We fix an integral symplectic basis $g^0,
g^a, g_a, g_0$ in $H^3(X, \complex)$. (This means that the vectors
of symplectic basis $g^A, g_A$ belong to the image of cohomology
with integral coefficients $H^3(X, \integers)$ in $H^3(X,
\complex)$. We use small Roman letters for indices running over
the set $\{ 1, 2, \cdots, r=h^{2,1}\}$ and capital Roman letters
for indices running over the set $\{0,1,\cdots, r=h^{2,1}\}$.)
Then special coordinates of $\widetilde{\calm}$ are defined by the
formula
\[
X^A=<g^A, \Omega>.
\]
Recall that $\dim_\complex \calm=h^{2,1},\ \dim_\complex
\widetilde{\calm}=\dim_\complex \calm +1=\frac{1}{2}\dim_\complex
H^3(X, \complex)$. Hence, we have the right number of coordinates.

The  functions $x^A=<g^A,\ >$ and $x_A=-<g_A,\ >$ define
symplectic coordinates on $H^3(X, \complex)$; on the lagrangian
submanifold $L$ we have
\[
x_A=\frac{\partial F_0(x^A)}{\partial x^A},
\]
where the function $F_0$ (the generating function of the
lagrangian submanifold $L$) has the physical meaning of genus 0
free energy. Notice that the lagrangian submanifold $L$ is
invariant respect to dilations (this is a consequence of the fact
that $\Omega$ is defined only up to a factor), and it follows that
$F_0$ is a homogeneous function of degree 2.

Identifying  $\widetilde{\calm}$  with the lagrangian submanifold
$L$ we see that the functions $x^A$ on $L$  are special
coordinates on $\widetilde{\calm}$ .

If two points of $\widetilde{\calm}$ correspond to the same point
of $\calm$ (to the same complex structure), then the forms
$\Omega$ are proportional; the same is true for the special
coordinates $X^A$. This means that $X^A$ can be regarded as
homogeneous coordinates on $\calm$. We can construct inhomogeneous
coordinates $t^1, \cdots, t^r$ by taking $t^i=\frac{X^i}{X^0},\
i=1, \cdots, h^{2,1}$. One can consider the free energy as a
function $f_0(t^1, \cdots, t^r)$; then
\[
F_0(X^0, \cdots, X^r)=(X^0)^2f_0(\frac{X^1}{X^0}, \cdots,
\frac{X^r}{X^0}).
\]

Let us work with the special coordinates $X^A=<g^A, \Omega>$ on
$\widetilde{\calm}$ and coordinates $t^a=\frac{X^a}{X^0}$ on
$\calm$. One can say that we are working with homogeneous
coordinates $X^0, \cdots, X^r$ assuming that $X^0=1$. We define
cohomology classes $\Omega_a$ on a Calabi-Yau manifold $X$ using
the formula
\[
\Omega_a=\gmconn_a \Omega+\omega_a \Omega,
\]
where $\omega_a$ are determined from the condition $\Omega_a\in
H^{2,1}$ and $\gmconn_a$ ($a=1, \cdots, r=h^{2,1}$) stands for the
Gauss-Manin covariant derivatives with respect to the special
coordinates $t^a=\frac{X^a}{X^0}$ on $\calm$. Representing
$\Omega$ as $X^A g_A+\partial_A F_0 g^A$ and taking into the
account that $g^A$ and $g_A$ are covariantly constant, we obtain
\[
\gmconn_a\Omega=g_a+\partial_a\partial_B F_0 g^B=g_A+\tau_{aB}g^B.
\]
(Recall that it follows from the Griffiths transversality that
$\gmconn_a\Omega\in H^{3,0}+H^{2,1}$. Every element of $H^{3,0}$
is represented in the form $\omega\Omega$; this follows from the
relation $H^{3,0}=\complex$.)

Now we can define a basis of $H^3(X, \complex)$ consisting of
vectors $(\Omega, \Omega_a, \overline{\Omega}_a,
\overline{\Omega})$. (It is obvious that $(\Omega, \Omega_a)$ span
$H^{3,0}+H^{2,1}$. Similarly, $\overline{\Omega}_a,
\overline{\Omega}$ span $H^{2,1}+H^{0,3}$.)  Let us introduce the
notation
\begin{equation}
e^{-K}=-i<\Omega, \overline{\Omega}>=i(\overline{X}^A
\frac{\partial F_0}{\partial X^A}-X^{A}\overline{\frac{\partial
F_0}{\partial X^A}}).
\end{equation}
(The function $K$ can be considered as a potential of a K\"ahler
metric on $\calm$.) Then we can calculate $\Omega_a$ using the
relation that $<\Omega_a, \overline{\Omega}>=0$; we obtain
\[
\Omega_a=\gmconn_a\Omega-\partial_aK\Omega.
\]

We can relate the basis $\{\Omega, \Omega_a, \overline{\Omega}_a,
\overline{\Omega}\}$ to the integral symplectic basis $\{g^A,
g_A\}$ by the following formulas
\begin{eqnarray}
\Omega=X^A g_A+\frac{\partial F_0}{\partial X^A}g^A,\ \ \ \
&\Omega_a=g_a+\frac{\partial^2 F_0}{\partial X^a\partial
X^B}g^B-\partial_aK(X^A g_A+\frac{\partial F_0}{\partial
X^A}g^A),\\
\overline{\Omega}=\overline{X}^A g_A+\overline{\frac{\partial
F_0}{\partial X^A}}g^A,\ \ \ \
&\overline{\Omega}_a=g_a+\overline{\frac{\partial^2 F_0}{\partial
X^a\partial X^B}}g^B-\partial_aK(\overline{X^A}
g_A+\overline{\frac{\partial F_0}{\partial X^A}}g^A).
\end{eqnarray}

The symplectic pairings between $\{\Omega, \Omega_i,
\overline{\Omega}_i, \overline{\Omega}\}$ are
\[
<\overline{\Omega}, \Omega>=-ie^{-K},\ \ \ \ \ \ \ <\overline{\Omega}_i,
\Omega_j>=-iG_{\bar{i}j}e^{-K},
\]
where $G_{i\bar{j}}$ is a K\"ahler metric on $\calm$ defined by
$G_{i\bar{j}}=\overline{\partial}_j\partial_iK$. As the
commutation relations are not the standard one, we introduce the
following cohomology classes
\[
\widetilde{\Omega}^i=iG^{i\bar{j}}e^K\overline{\Omega}_j,\ \ \ \
\widetilde{\Omega}=ie^K\overline{\Omega}.
\]
Due to the relations
\[
<\widetilde{\Omega}, \Omega >=1,\ \ \ \ \ \ \ <\widetilde{\Omega}^i, \Omega_j
>=\delta_i^j,
\]
we can say that $\{\Omega, \Omega_a, \widetilde{\Omega}^a,
\widetilde{\Omega}\}$ constitutes a symplectic basis, which
specifies a complex hermitian polarization.

Directly differentiating the above expressions with respect to the
parameters  $t^a$ and $\bar{t}^a$, we have
\begin{equation}
\label{eq:hol-trans}
\begin{split}
\gmconn_i\Omega=\Omega_i-\partial_i K \Omega,\ \ \ \ &
\gmconn_i\Omega_j=-\partial_iK\Omega_j+\Gamma_{ij}^k\Omega_k+iC_{ijk}\widetilde{\Omega}^k,\\
\gmconn_i\widetilde{\Omega}=\partial_iK \widetilde{\Omega},\ \ \ \
&\gmconn_i\widetilde{\Omega}^j=\partial_iK
\widetilde{\Omega}^j-\sum_k\Gamma_{ij}^k\widetilde{\Omega}^k-\widetilde{\Omega}\delta_{ij};
\end{split}
\end{equation}
where $\Gamma_{ij}^k$ is the Christoeffel symbol for the K\"ahler metric
$G_{i\bar{j}}$. And
\begin{equation}
\label{eq:anti-holo-trans}
\begin{split}
\overline{\gmconn}_i\Omega=0,\ \ \ \ \
&\overline{\gmconn}_i\Omega_j=G_{\bar{i}j}\Omega,\\
\overline{\gmconn}_i\widetilde{\Omega}=-G_{\bar{i}j}\widetilde{\Omega}^j,\
\ \ \ \
&\overline{\gmconn}_i\widetilde{\Omega}^j=-ie^{2K}\overline{C}_{\bar{i}}^{jk}\Omega_k.
\end{split}
\end{equation}
Applying (\ref{eq:Z}), we obtain from (\ref{eq:hol-trans}),
(\ref{eq:anti-holo-trans}) equations governing the dependence of
the state $\Psi(x^I, t^i, \bar{t}^i, \lambda)$ on $t^i,
\overline{t}^i$ (holomorphic anomaly equation).
\begin{eqnarray}\label{eq:holom-anomaly}
\frac{\partial \Psi}{\partial
\bar{t}^i}&=\big[\frac{1}{2}\lambda^2e^{2K}\overline{C}_{\bar{i}\bar{j}\bar{k}}G^{\bar{j}j}G^{\bar{k}k}\frac{\partial^2}{\partial
x^i\partial x^j}+G_{\bar{i}j}x^j\frac{\partial}{\partial x^0}+C_i\big]\Psi,\\
\frac{\partial \Psi}{\partial t^i}&=[x^0\frac{\partial}{\partial
x^i}-\partial_iK(x^0\frac{\partial}{\partial
x^0}+x^j\frac{\partial}{\partial
x^j})-\Gamma_{ij}^kx^j\frac{\partial}{\partial
x^k}-\frac{1}{2}\lambda^{-2}C_{ijk}x^jx^k+D_i]\Psi.
\end{eqnarray}

Notice that usually these equations are written with
$C_i=0, D_i=0.$ This is possible if we consider only one of these equations; fixing $C_i$ or $D_i$ corresponds to (physically
irrelevant) choice of normalization of the wave function.
However, in general it is impossible to  assume that
$C_i=0, D_i=0.$  (We can eliminate $C_i$ or $D_i$ changing the normalization of the wave function, but we cannot eliminate both of them.) Let us emphasize that $C_i$ and $D_i$ are constraint by the requirement that (\ref {eq:holom-anomaly}) has a solution.

\section{Holomorphic polarization}
Let us start again with the integral symplectic basis $\{g_0,
g_a,g^a, g^0\}$. We will normalize the form $\Omega$ requiring
that $<g^0, \Omega>=X^0=1$. We would like to define a symplectic
basis in the middle dimensional cohomology that depends
holomorphically on the points of moduli space. Namely, we will
consider the following basis in $H^3(X, \complex)$,
\[
\begin{split}
e^0&=g^0,\\
e^a&=g^a-t^ag^0,\\
e_a&=\gmconn_a\Omega,\\
e_0&=\Omega=g_0+X^ag_a+\partial_a F_0g^a+\partial_0F_0g^0,
\end{split}
\]
where $\gmconn_a$ stands for the Gauss-Manin connection. It is easy to check that this basis is symplectic.

Using the above relations, we obtain an expression of the new
basis in terms of the integral symplectic basis $g^A, g_A$,
\begin{equation}
\label{eq:hol-basis-real}
\begin{array}{lll}
&e^0&=g^0\\
&e^a&=g^a-t^ag^0,\\
&e_0&=g_0+t^ag_a+\frac{\partial f_0}{\partial t^a}g^a+
(2f_0-t^a\frac{\partial f_0}{\partial t^a})g^0,\\
&e_a&=g_a+\frac{\partial ^2 f_0}{\partial t^a\partial
t^b}g^b+(\frac{\partial f_0}{\partial t^a}-t^b\frac{\partial^2
f_0}{\partial t^a\partial t^b})g^0.
\end{array}
\end{equation}
Let $q_A, q^A$ denote the coordinates on the basis $g^A, g_A$, and
$\epsilon_A, \epsilon^A$ the coordinates on the basis $e^A, e_A$.
Equation (\ref{eq:r}) permits us to relate the wave function  in
real polarization to the wave function in our new basis (in
holomorphic polarization)
\begin{equation}\label{eq:trans-hol-real}
\Psi_{\text {hol}}(\epsilon^I, t^i,
\lambda)=e^{-\frac{1}{2}\lambda^{-2}R_{AB} \epsilon^A
\epsilon^B}\Psi_{\text {real}}(\epsilon^0,
\epsilon^i+t^i\epsilon^0, \lambda),
\end{equation}
where $R$ is a the matrix
\[
\left(\begin{array}{cc}2f_0
& \frac{\partial f_0}{\partial t^a}\\
\frac{\partial f_0}{\partial t^a}&\frac{\partial^2f_0}{\partial
t^a\partial t^b}\end{array}\right).
\]
Notice that $\Psi_{\text{hol}}$ is defined up to a t-dependent
factor; we use Equation (\ref{eq:trans-hol-real}) to fix this
factor.

Using that $g^0, g^a, g_a, g_0$ are covariantly constant with
respect to the Gauss-Manin connection $\gmconn_a$, we see that
\begin{equation}\label{eq:gauss-manin-conn}
\begin{array}{lll}
&\gmconn_b e_0=e_b,\ \ \ \ \ \ & \gmconn_b e_a=C_{abc} e^c\\
&\gmconn_b e^a=\delta_{ab}e^0,\ \ \ \ \ \ & \gmconn_b e^0=0,
\end{array}
\end{equation}
where $C_{abc}=\partial_a\partial_b\partial_c f_0$. Applying
Equation (\ref{eq:Z}), we obtain from Equation
(\ref{eq:gauss-manin-conn}) the dependence of the state
$\Psi_{\text {hol}}(\epsilon^A, t^i, \lambda)$ on the coordinates
$t^1, \cdots, t^h$
\begin{equation}
\label{eq:holom-anomaly-hol} \frac{\partial
\Psi_{\text{hol}}(\epsilon^A, t^i, \lambda)}{\partial
t^a}=(\epsilon^0\frac{\partial}{\partial
\epsilon^a}-\frac{1}{2}\lambda^{-2}C_{abc}\epsilon^b\epsilon^c+\sigma_a(t))\Psi_{\text{hol}}(\epsilon^A,
t^i, \lambda).
\end{equation}
The function $\Psi_{\text{hol}}$ defined by the Equation
(\ref{eq:trans-hol-real}) obeys Equation
 (\ref{eq:holom-anomaly-hol}) with $\sigma_a=0$.
We remark that because our basis is holomorphic, the state $\Psi$
does not depend on antiholomorophic variables $\bar{t}^i$.
Therefore Equation (\ref{eq:holom-anomaly-hol}) is the only
equation the state $\Psi_{\text{hol}}(\epsilon_i, t^i, \lambda)$
has to satisfy. This equation can be easily solved. The solution
can be written as follows,
\begin{equation}
\begin{array}{lll}
&\Psi&=\exp(W_1+W_2),\\
&W_1&=W(\epsilon^0, \epsilon^0t^a+\epsilon^a),\\
&W_2&=-\lambda^{-2}(\frac{1}{2}\frac{\partial^a f_0}{\partial
t^i\partial t^j}\epsilon^i\epsilon^j+\frac{\partial f_0}{\partial
t^i}\epsilon^i\epsilon^0+f_0(\epsilon^0)^2),
\end{array}
\end{equation}
where $W$ is an arbitrary function of $h^{2,1}+1$ variables.

Comparing the above expression with Equation
(\ref{eq:trans-hol-real}), we obtain
\begin{equation}
\exp(W)=\Psi_{\text {real}}.
\end{equation}

Let us consider now the B-model in the neighborhood of the
maximally unipotent boundary point. We choose $g^0$ as covariantly
constant cohomology class that can be extended to the boundary
point and we define $g^a$ as covariantly constant cohomology
classes having logarithmic singularities at the boundary point.
The special coordinates coincide with the canonical coordinates
and the basis $\{e^A, e_A\}$ coincides with the basis that is
widely used in the theory of mirror symmetry. (See
\cite{cox-katz:book}, Section 6.3). This can be derived, for
example, from the fact that the Gauss-Manin connection described
by the formula (\ref {eq:gauss-manin-conn}
) has the same form in both bases.

\section{Partition function of B-model}
The partition function $\Psi$ of the topological sigma model on a
Calabi-Yau threefold $X$ (and more generally of twisted $N=2$
superconformal theory) can be represented as $\Psi=e^F$, where
\begin{equation}
\label{eq:F} F=\sum_g\lambda^{2g-2}F_g(t,\bar{t}),
\end{equation}
and $F_g$ has a meaning of contribution of surfaces of genus $g$
to the free energy. The correlation functions $C_{i_1,\cdots, i_n}
^{(g)}$ can be obtained from $F_g$ by means of covariant
differentiation. Notice that in Equation (\ref{eq:F}) we can
consider $t$ and $\bar{t}$ as independent complex variables. The
covariant derivatives with respect to $t$ coincide  with
$\frac{\partial}{\partial t^i}$ in the limit when $\bar{t}\to
\infty$ and $t$ remains finite.

It is convenient to introduce the generating functional of
correlation functions
\begin{equation}\label{eq:gen-func}
W(\lambda, x, t, \bar{t})=\sum_{g=0}^\infty\sum_{n=1}^\infty
\frac{1}{n!}\lambda^{2g-2}C^{(g)}_{i_1,\cdots, i_n}x^{i_1}\cdots
x^{i_n}+(\frac{\chi}{24}-1)\log(\lambda),
\end{equation}
where $C^{(g)}_{i_1,\cdots, i_n}=0$ for $2g-2+n\leq 0$. The number
$\chi$ is defined as the difference between the numbers of the
bosonic and fermionic modes; in the case of topological
sigma-model it coincides with the Euler characteristic of $X$ (up
to a sign).

The function $W$ obeys the following holomorphic anomaly equations
(Equation 3.17, 3.18, \cite{BCOV:hol-anomaly})
\begin{equation}
\label{eq:holom-anomaly-1} \frac{\partial}{\partial
\bar{t}^i}\exp(W)=\big[\frac{\lambda^2}{2}
\overline{C}_{ijk}e^{2K}G^{j\bar{j}}G^{k\bar{k}}\frac{\partial^2}{\partial
x^j\partial x^k} -G_{\bar{i}j}x^j(\lambda\frac{\partial}{\partial
\lambda}+x^k\frac{\partial}{\partial x^k})\big]\exp(W),
\end{equation}
and
\begin{equation}
\label{eq:holom-anomaly-2}
\begin{split}
&\big[\frac{\partial}{\partial
t^i}+\Gamma^k_{ij}x^j\frac{\partial}{\partial
x^k}+\partial_iK(\frac{\chi}{24}-1-
\lambda\frac{\partial}{\partial \lambda})\big]\exp(W)\\
&=\big[\frac{\partial}{\partial x^i}-\partial_iF_1-
\frac{1}{2\lambda^2}C_{ijk}x^jx^k\big]\exp(W).
\end{split}
\end{equation}
One can modify the definition of $W$ by introducing a new function
$\widetilde{W} $,
\begin{equation}
\begin{split}
&\widetilde{W}(\lambda, x^i, \rho,
t,\bar{t})=\sum\limits_{g=1}^\infty\sum\limits_{n=1}^\infty\frac{1}{n!}\lambda^{2g-2}
C_{i_1,\cdots, i_n}^{(g)}x^{i_1}\cdots x^{i_n}\rho^{-n-(2g-2)}+\nonumber\\
&+(\frac{\chi}{24}-1)\log\rho=W(\frac{\lambda}{\rho},
\frac{x}{\rho}, t,\bar{t})-(\frac{\chi}{24}-1)\log(\lambda).
\end{split}
\end{equation}
The function $\widetilde{W}$ (we will call it BCOV wave function) satisfies the equations
\begin{equation}
\label{eq:holo-anomal-twisted}
\begin{split}
&\frac{\partial}{\partial
\bar{t}^i}\exp(\widetilde{W})=
\big[\frac{\lambda^2}{2}\overline{C}_{\bar{i}}^{jk}\frac{\partial^2}{\partial
x^j\partial x^k}
+G_{\bar{i}j}x^j\frac{\partial}{\partial \rho}\big]\exp(\widetilde{W}),
\end{split}
\end{equation}
\begin{equation}
\label{eq:anti-holo-anomal-twisted}
\begin{split}
\frac{\partial}{\partial
t^i}\exp(\widetilde{W})&=\big[\rho\frac{\partial}{\partial x^i}
-\partial_iK(\rho\frac{\partial}{\partial
\rho}+x^j\frac{\partial}{\partial
x^j})-\Gamma_{ij}^kx^j\frac{\partial}{\partial x^k}
-\frac{1}{2\lambda^2}C_{ajk}x^jx^k\\
&-\partial_iF_1-\partial_iK(\frac{\chi}{24}-1)\big]\exp(\widetilde{W}).
\end{split}
\end{equation}
Equations (\ref{eq:holo-anomal-twisted}) follows from
(\ref{eq:holom-anomaly-1}); it is equivalent to the Equation
(6.11) of \cite{BCOV:hol-anomaly}. And Equation (\ref{eq:anti-holo-anomal-twisted}) follows
from Equation (\ref{eq:holom-anomaly-2}) .

The above equations are valid for any topologically twisted $N=2$
superconformal theory coupled to gravity. We will apply them to
the study of B-model. In this case, it is clear from comparison of
Equations (\ref{eq:holo-anomal-twisted}) with
(\ref{eq:holom-anomaly}) that $\exp(\widetilde{W})$ can be
interpreted as a wave function corresponding to the complex
hermitian polarization considered in Section 3. The Equation
(\ref{eq:holo-anomal-twisted}) implies that $\exp(\widetilde{W})$
is a projectively flat section. Let us emphasize that
$\exp(\widetilde{W})$ is a well defined function determined from
worldsheet considerations and the wave function is specified only
up to a factor.  Considering  $\exp(\widetilde{W})$  as a wave
function we fix this factor. The fact that the wave function in
complex hermitian polarization can be considered as a one-valued
function on the whole moduli space of complex structures
(monodromy transformations act trivially) was essentially used in
\cite{abk:modular}. Notice also that the worldsheet interpretation
of the wave function permits us to analyze its behavior at
boundary points of the moduli space; this information imposes
further restrictions on the quantum state obtained by quantization
of genus zero theory.

The function $\widetilde{W}(\lambda, x, \rho, t, \infty)$ can be
represented in terms of $F(\lambda, t, \infty)$
 in the following way
\begin{eqnarray}\label{eq:w-tilde}
\widetilde{W}(\lambda, x, \rho, t,
\infty)=&\sum_g\frac{1}{n!}(\frac{\lambda}{\rho})^{2g-2}F_g(t+\frac{x}{\rho},
\infty)
-F_1(t, \infty)
-(\frac{\lambda}{\rho})^{-2}(F_0(t, \infty)\nonumber\\
&+\frac{\partial F_0(t, \infty)}{\partial t^i}\cdot
\frac{x^i}{\rho}
+\frac{1}{2}\frac{\partial^2 F_0(t, \infty)}{\partial t^i\partial t^j}
\cdot \frac{x^ix^j}{\rho^2})-(\frac{\chi}{24}-1)\log\rho\nonumber\\
=&F(\frac{\lambda}{\rho}, t+\frac{x}{\rho}, \infty)-F_1(t, \infty)
-(\frac{\lambda}{\rho})^{-2}(F_0(t, \infty)+\nonumber\\
&\frac{\partial F_0(t, \infty)}{\partial t^i}\cdot
\frac{x^i}{\rho} +\frac{1}{2}\frac{\partial^2 F_0(t,
\infty)}{\partial t^i\partial t^j}\cdot
\frac{x^ix^j}{\rho^2})-(\frac{\chi}{24}-1)\log\rho\
 \end{eqnarray}
(Due to the relation $C_{i_1, \cdots,
i_n}^{(g)}=\partial_{i_1}\cdots \partial _{i_n}F_g(\lambda, t,
\infty)$, the expression for $\widetilde{W}$ as $\bar{t}\to
\infty$ can be considered as Taylor series; for $g=0$ and $g=1$
the first few terms
 of the Taylor series are missing because one assumes that $C^{(g)}_{i_1, \cdots, i_n}=0$ for $2g-2+n\leq 0$.)
 Notice that the above formula can be used both for A-model and for B-model (in the latter case $t^a$ are
 canonical coordinates in the neighborhood of maximally unipotent boundary point).

As we have seen  in the language of B-model,
$\exp(\widetilde{W}(\lambda, x, \rho, t, \infty))$ can be
interpreted as a wave function in the complex hermitian
polarization for $\bar{t}= \infty$. From the other side, the
complex hermitian polarization for $\bar{t}=\infty$ coincides with
holomorphic polarization (see Appendix). This means that up to a
$t$-dependent factor $\exp(\widetilde{W}(\lambda, x, \rho, t,
\infty))$ coincides with the wave function in holomorphic
polarization.

Let us give another proof of this fact that permits us to
calculate the $t$-dependent factor.

We defined the wave function in holomorphic polarization as a
solution to the equation (\ref{eq:holom-anomaly-hol}). The
function $\exp(\widetilde{W}(\lambda, x, \rho, t, \infty))$ obeys
a little bit different equation (for $\bar{t}=\infty$, we can take
$\partial_aK=0$, $\Gamma_{aj}^k=0$ in the equation
(\ref{eq:holo-anomal-twisted})). Comparing these equations, we see
that (up to a constant factor)
\begin{equation}
\label{eq:psi} \Psi(\lambda, x, \rho,
t)=\exp(\widetilde{W}(\lambda, x, t,\rho, \infty))\exp(F_1).
 \end{equation}
 It follows from this expression that our function $ \Psi(\lambda, x, \rho,t)$ coincides with the modification of  BCOV wave function
 $\exp(\widetilde{W}(\lambda, x, t,\rho, \infty))$ considered in [11], [6].

Using the expression (\ref{eq:w-tilde}) of $\widetilde{W}$ and
(\ref {eq:psi}), we have the following expression for $\Psi$
\begin{eqnarray}
\Psi&=\exp\big(F(\frac{\lambda}{\rho}, t+\frac{x}{\rho},
\infty)-\lambda^{-2}(F_0(t,\infty)\rho^2+\frac{\partial
F_0}{\partial t^i}x^i\rho+\frac{1}{2}\frac{\partial^2
F_0}{\partial t^i\partial
t^j}x^ix^j)\nonumber\\
&-(\frac{\chi}{24}-1)\log(\rho)\big).
\end{eqnarray}
We use Equation (\ref{eq:trans-hol-real}) to compute the
corresponding wave function in the real polarization by
identifying $\rho=\epsilon^0, x^i=\epsilon^i$.
\begin{equation}
\Psi_{\text{real}}(\epsilon^0, \epsilon^i+t^i\epsilon^0,
\lambda)=\exp(F(\frac{\lambda}{\epsilon^0},
t^i+\frac{\epsilon^i}{\epsilon^0},
\infty)-(\frac{\chi}{24}-1)\log(\epsilon^0)).
\end{equation}
Accordingly, if we set $x^0=\epsilon^0$ and
$x^i=\epsilon^i+t^i\epsilon^0$, we
have
\begin{equation}
\Psi_{\text{real}}(x^0, x^i, \lambda)=\exp(F(\frac{\lambda}{x^0}, \frac{x^i}{x^0},
\infty)-(\frac{\chi}{24}-1)\log(x^0)).
\end{equation}

It follows from this equation that
\begin{equation}
\Psi_{\text{real}}(cx^0, cx^i, c\lambda)= \Psi_{\text{real}}(x^0, x^i, \lambda)c^{-(\frac{\chi}{24}-1)}
\end{equation}

Using (10) one can conclude that similar homogeneity property
is valid in any polarization:
\begin{equation}
\Psi (cx, e, c\lambda)= \Psi (x, e, \lambda)c^{-(\frac{\chi}{24}-1)}
\end{equation}

To clarify the physical meaning of $\exp(\widetilde{W})$, it is
convenient to consider the mirror A-model and to take the limit
$\bar{t}\to \infty$. Then the free energy $F_g$ and therefore the
functions $W$ and $\widetilde{W}$ can be expressed in terms of
Gopakumar-Vafa invariants $n_\beta^g$ and topological invariants
of the mirror manifold $\widetilde {X}$. Namely
\begin{equation}
\label {F}
F=ln\Psi=\sum\limits_g\lambda^{2g-2}F_g(t)=F'+F''
\end{equation}
can be represented as a sum of two summands $F'$ and $F''$ where
$F'$ corresponds to non-trivial instanton contribution of the
mirror A-model with the form
\begin{equation}
\label{F'}
F'=\sum_{n,g,\beta\ne0}n_\beta^g \frac{1}{m}(2\sin
\frac{m\lambda}{2})^{2g-2}e^{nt^\beta},
\end{equation}
and the constant map contribution $F''$ can be represented as
\begin{equation}
\label{F''}
F''=const+\lambda^{-2}\sum\frac{t^{\beta_1}t^{\beta_2}t^{\beta_3}}{3!}\int_{\widetilde {X}}
\beta_1\cup \beta_2\cup \beta_3-\frac{t^\beta}{24}\int_{\widetilde {X}}\beta\cup
c_2({\widetilde {X}}).
\end{equation}
In (\ref{F'}) we assume that $\beta$ runs over the two-dimensional integral
homology group of $\widetilde {X}$ (more precisely, only the elements in the positive cone of this group are relevant); in (\ref{F''}) $\beta$ runs over a basis of this group. Recall that the two-dimensional cohomology group labels the deformations of
K\"ahler structures on $\widetilde {X}$; in the language of mirror B-model it
corresponds to the
cohomology group $H^{2,1}(X)$ that labels deformations of
complex  structures; the coordinates $t^{\beta}$ correspond
to canonical coordinates on the moduli space of complex structures
of the corresponding B-model.

Instead of free energy $F=F'+F''$ one can consider the partition
function $Z=e^F$ represented as a product of two factors
$Z'=e^{F'}$ and $Z''=e^{F''}.$ By means of formal manipulations
(see \cite{katz:gopakumar}), one can derive from (\ref {F'}) the
following expression\footnote {To give a precise meaning to (\ref
{Z'}) one can consider  this expression as an element of Novikov
ring with generators $q^{\beta}$ and with coefficients in Laurent
series with respect to $\Lambda$; the generators obey the relation
$ q^{\beta}q^{\beta '}= q^{\beta +\beta '}$. } for $Z':$
\begin{equation}
\label{Z'} Z'=\prod (1-\Lambda ^s q^{\beta})^{m^s_{\beta}}
\end{equation}
where
\begin{eqnarray}
m_\beta^s&=&sn_\beta^0+(-1)^{1+s}\sum_{g\geq1+|s|}n_\beta^g\left(\begin{array}{l}2g-2\\
g-1-s\end{array}\right),\\
\Lambda&=&e^{-i\lambda},\\
q^{\beta}&=&\exp {t^{\beta}}.
\end{eqnarray}

Using the expression (\ref{eq:psi}), we obtain an expression of
the wave function in holomorphic polarization:
\[
\Psi(\lambda, x, \rho, t)=\Psi^{'}\Psi^{''},
\]
where $\Psi'$ is expressed in terms of Gopakumar-Vafa invariants
$n_\beta^g$ with $g\geq 0$. More precisely,
\begin{eqnarray}
\Psi'(\lambda, x, \rho, t)=&\exp(\sum_{m,g\geq1, \beta\ne 0}n_\beta^g
\frac{1}{m}(2\sin \frac{m\lambda}{2\rho})^{2g-2}e^{m(t^\beta+\frac{x^\beta}{\rho})})\nonumber\\
=&\prod_{s,\beta\ne
0}(1-e^{-is\frac{\lambda}{\rho}+(t^\beta+\frac{x^\beta}{\rho})})^{m_\beta^s}.
\end{eqnarray}
\vskip 5mm

\noindent{\Large \bf Appendix: Relation between complex hermitian
polarization and holomorphic polarization}\vskip 2mm

\noindent{Here}  we relate the holomorphic polarization to the
complex hermitian polarization.

Since both bases are expressed in terms of the real basis $\{g_A,
g^A\}$, we can compute the expression of $\{\Omega, \Omega_a,
\widetilde{\Omega}^a, \widetilde{\Omega}\}$ by $\{ e_0, e_a, e^a,
e^0\}$ as follows.
\begin{equation}
\label{eq:trans-hol-herm} \begin{array}{llll}
&\Omega&=e_0,\\ &\Omega_a&=e_a-\partial_aK e_0,\\
&\widetilde{\Omega}&=ie^Ke_0+ie^K(\overline{X}^a-X^a)e_a+\partial_aKe^a+e^0,\\
&\widetilde{\Omega}^a&=G^{a\bar{b}}\big[-ie^K\overline{\partial_bK}e_0+ie^Ke_b-ie^K(\overline{X}^c-X^c)
\overline{\partial_bK}e_c\\
&&+(G_{\bar{b}c}+\partial_bK\overline{\partial_cK}-\partial_cK\overline{\partial_bK})e^c\big]
\end{array}
\end{equation}

We recall that in the neighborhood of maximally unipotent boundary point, the function $F_0$ has the
following expression
\[
F_0(X^0, X^i)=\frac{d_{ijk}X^iX^jX^k}{X^0}+c(X^0)^2+\sigma(q^i),
\]
where $\sigma$ is a holomorphic function and $q^j\stackrel{def}{=}\exp(2\pi i t^j)$.  The expression  $\sigma(q^i)$ is bounded in the neighborhood of the boundary point $ q^i=0.$

In the following computation, we set $X^0=1$. Substituting the
above expression of $F_0$ into $K=-\log(i(\overline{X}^A
\frac{\partial F_0}{\partial X^A}-X^A\overline{\frac{\partial
F_0}{\partial X^A}}))$, we obtain
\[
K=-\log\big(i\big(\bar{d}_{ijk}\bar{t}^i\bar{t}^j
\bar{t}^j-d_{ijk}t^it^j t^k+3d_{ijk}\bar{t}^it^jt^k
-3\bar{d}_{ijk}t^i\bar{t}^j\bar{t}^k+2(c-\bar{c})+\varphi\big)
\big),
\]
where $\varphi$ is equal to
\[
-\log(q^i)q^i\frac{\partial \sigma}{\partial
q^i}+\log(\bar{q}^i)q^i\frac{\partial \sigma}{\partial
q^i}+\log(\bar{q}^i)\bar{q}^i\overline{\frac{\partial
\sigma}{\partial q^i}}-\log(q^i)\bar{q}^i\overline{\frac{\partial
\sigma}{\partial q^i}}.
\]

Differentiating $K$ respect to $t^a$, we have
\[
\partial_aK=\frac{3d_{ajk}t^jt^k-6d_{aik}\bar{t}^it^k
+3\bar{d}_{ajk}\bar{t}^j\bar{t}^k-\partial_a\varphi}{\bar{d}_{ijk}\bar{t}^i\bar{t}^j
\bar{t}^j-d_{ijk}t^it^j t^k+3d_{ijk}\bar{t}^it^jt^k
-3\bar{d}_{ijk}t^i\bar{t}^j\bar{t}^k+2(c-\bar{c})+\varphi}
\]
Similarly,
\begin{equation}
\overline{\partial_aK}=\frac{3\bar{d}_{ajk}\bar{t}^j\bar{t}^k-
6\bar{d}_{aik}t^i\bar{t}^k
+3d_{ajk}t^jt^k-\overline{\partial_a\varphi}}{d_{ijk}t^it^j
t^j-\bar{d}_{ijk}\bar{t}^i\bar{t}^j
\bar{t}^k+3\bar{d}_{ijk}t^i\bar{t}^j\bar{t}^k -3d_{ijk}\bar{t}^i
t^j t^k+2(\bar{c}-c)+\overline{\varphi}}.
\end{equation}

Taking the derivative of $\partial_aK$ respect to $\bar{t}^b$,
we obtain
\begin{equation}
\begin{split}
\bar{\partial}_b\partial_aK= \frac{
    -6d_{bak}t^k+6\bar{d}_{abk}\bar{t}^k}
{\bar{d}_{ijk}\bar{t}^i\bar{t}^j \bar{t}^j-d_{ijk}t^it^j
t^k+3d_{ijk}\bar{t}^it^jt^k
-3\bar{d}_{ijk}t^i\bar{t}^j\bar{t}^k+2(c-\bar{c})+\varphi}+\\
-\frac{\big( 3d_{ajk}t^jt^k-6d_{aik}\bar{t}^it^k
+3\bar{d}_{ajk}\bar{t}^j\bar{t}^k+\partial_a\varphi\big)\big(3\bar{d}_{bjk}\bar{t}^j\bar{t}^k-
6\bar{d}_{bik}t^i\bar{t}^k +3d_{bjk}t^jt^k\big)}
{\big(\bar{d}_{ijk}\bar{t}^i\bar{t}^j \bar{t}^j-d_{ijk}t^it^j
t^k+3d_{ijk}\bar{t}^it^jt^k
-3\bar{d}_{ijk}t^i\bar{t}^j\bar{t}^k+2(c-\bar{c})+\varphi\big)^2}
\end{split}
\end{equation}

Also we have
\[
e^{K}=\frac{1}{e^{-K}}=\frac{1}{i\big(\bar{d}_{ijk}\bar{t}^i\bar{t}^j
\bar{t}^j-d_{ijk}t^it^j t^k+3d_{ijk}\bar{t}^it^jt^k
-3\bar{d}_{ijk}t^i\bar{t}^j\bar{t}^k+2(c-\bar{c})+\varphi\big)}.
\]

Let us consider $t$ and $\bar{t}$ in these formulas as independent
complex variables; then the basis $(\Omega, \Omega_a,
\widetilde{\Omega}^a, \widetilde{\Omega})$ is not hermitian anymore. We will check that this basis tends to $(e_0, e_a, e^a,
e^0)$ as  $\bar{q}$ converges to 0 (and therefore $\bar{t}\rightarrow \infty$)  with fixed $t$.

  Fixing
$t$ and taking $\bar{t}\to \infty$, we have that
\[
\varphi\sim O(\bar{t}),\
\partial_a\varphi\sim O(\bar{t}),\ \overline{\partial_a\varphi}\sim
O(1),
\]
where $O(1)$ stands for bounded terms.

And therefore, we have the following asymptotic leading terms.
\begin{enumerate}
\item
$\partial_aK\sim3\frac{\bar{d}_{ajk}\overline{t}^j\overline{t}^k}{\bar{d}_{ijk}\overline{t}^i\overline{t}^j\overline{t}^k}$;
\item
$\overline{\partial_a
K}\sim-3\frac{\bar{d}_{ajk}\overline{t}^j\overline{t}^k}{\bar{d}_{ijk}\overline{t}^i\overline{t}^j\overline{t}^k}$;
\item
$G_{a\bar{b}}\sim
\frac{6\bar{d}_{abk}\overline{t}^k}{\bar{d}_{ijk}\overline{t}^i\overline{t}^j\overline{t}^k}-\frac{9\bar{d}_{ajk}\bar{d}_{blm}\overline{t}^j\overline{t}^k\overline{t}^l\overline{t}^m}
{(\bar{d}_{ijk}\overline{t}^i\overline{t}^j\overline{t}^k)^2}$
\item
$e^K\sim
\frac{1}{i\bar{d}_{ijk}\overline{t}^i\overline{t}^j\overline{t}^k}$;
\item
$\partial_aK\overline{\partial_bK}-\partial_bK\overline{\partial_aK}\sim\frac
{\left(\begin{array}{l}d_{iaj}\bar{d}_{bkl}\overline{X}^i\overline{t}^k\overline{t}^lt^j+\bar{d}_{aij}\bar{d}_{kbl}
\overline{t}^i\overline{t}^j\overline{t}^lt^k\\
-d_{ibj}\bar{d}_{akl}\overline{t}^i\overline{t}^k\overline{t}^lt^j
-\bar{d}_{bij}\bar{d}_{kal}\overline{t}^i\overline{t}^j\overline{t}^lt^k\end{array}\right)}
{\bar{d}_{ijk}d_{lmn}\overline{t}^i\overline{t}^j\overline{t}^k\overline{t}^l\overline{t}^m\overline{t}^n}
$.
\end{enumerate}

If we let $\bar{t}^i=\bar{s}^i\nu$ and $\nu\to \infty$, we have
the following asymptotic behavior (up to a factor),
\begin{enumerate}
\item $\partial_aK\sim \frac{1}{\nu}$;
\item $\overline{\partial_aK}\sim \frac{1}{\nu}$;
\item $G_{a\bar{b}}\sim \frac{1}{\lambda^2}L_{a\bar{b}}$, where $L_{a\bar{b}}$ is a nondegenerate anti-holomorphic matrix;
\item $e^K\sim \frac{1}{\nu^3}$;
\item $\partial_aK\overline{\partial_bK}-\partial_bK\overline{\partial_aK}\sim
\frac{1}{\nu^3}$.
\end{enumerate}

Substituting the above asymptotic expressions  into Equation
(\ref{eq:trans-hol-herm}) and taking the limit $\bar{t}\to \infty$ by
letting $\nu\to \infty$, we have
\begin{equation}
\begin{array}{lll}
&\Omega=e_0,\ \ \ \ \ \ &\Omega_a=e_a,\\
&\widetilde{\Omega}=e^0,\ \ \ \ \ \ &\widetilde{\Omega}^a=e^a.
\end{array}
\end{equation}

\vskip 2mm

\noindent{Albert Schwarz}

\noindent{Department of Mathematics}, University of California,
Davis, CA, 95616, U.S.A.

\noindent{schwarz@math.ucdavis.edu}\\

\noindent{Xiang Tang}

\noindent{Department of Mathematics}, Washington University, St.
Louis, MO, 63130, U.S.A.

\noindent{xtang@math.wustl.edu}

\begin{thebibliography}{99}
\bibitem{abk:modular}
Aganagic, M., Bouchard, V., Klemm, A.,Topological Strings and
(Almost) Modular Forms, {\em arxiv:hep-th/0607100}.

\bibitem{BCOV:hol-anomaly}
Bershadsky, M., Cecotti, S., Ooguri, H., and Vafa, C.,
Kodaira-Spencer Theory of Gravity and Exact Results for Quantum
String Amplitude, {\em Comm. Math. Phys.}, 165 (1994), 211-408.

\bibitem{cox-katz:book}
Cox, D., and Katz, S., {\em Mirror symmetry and algebraic
geometry}, Mathematical Surveys and Monographs, 68. American
Mathematical Society, Providence, RI, 1999.

\bibitem{de:metaplectic}
Deligne, P., Groupe de Heisenberg et r\'ealit\'e, {\em J. Amer. Math.
Soc.}, 4 (1991), no. 1, 197--206.

\bibitem{gs:quantization}
Gerasimov, A., Shatashvili, S.,  Towards integrability of topological strings. I. Three-forms on Calabi-Yau manifolds, {\em J. High Energy Phys.} 2004, no. 11.

\bibitem{gnp}
Gunaydin,M., Neitzke,A.,  Pioline,B.
 Topological wave functions and heat equations,{\em
arxiv:hep-th/0607200}

\bibitem{katz:gopakumar}
Katz, S., Gromov-Witten, Gopakumar-Vafa, and Donaldson-Thomas
invariants of Calabi-Yau threefolds, {\em arxiv: math.AG/0408266}.

\bibitem{ksv:integrality}
Kontsevich, M., Schwarz, A., and Vologodsky, V., Integrality of
instanton numbers and $p$-adic B-model,  {\em Phys. Lett. B  637}
(2006),  no. 1-2, 97--101.

\bibitem{Loran:quantization}
Loran, F., K\"ahler quantization of $H\sp 3({\rm CY}\sb 3,R)$ and the holomorphic anomaly,  {\em J. High Energy Phys.},  2005, no. 12, 004.

\bibitem{sv}
Schwarz, A., Vologodsky, V. Frobenius transformation, mirror map and instanton numbers {\em arxiv:hep-th/0606151}

\bibitem{verlinde:attractor}
Verlinde, E., Attractors and the Holomorphic Anomaly, {\em
arxiv:hep-th/0412139}.

\bibitem{witten:hol-anomaly}
Witten, E., Quantum Background Independence In String Theory, {\em
arxiv:hep-th/9306112}.

\bibitem{yy:partition}
Yamaguchi, S., Yau, S.T., Topological string partition functions
as polynomials,  {\em J. High Energy Phys.} 2004, no. 7, 047,
\end{thebibliography}
\end{document}